\documentclass[aps,pra,twocolumn,superscriptaddress,nofootinbib,10pt]{revtex4-2}

\usepackage[T1]{fontenc}

\usepackage{amsmath,amssymb,bm,mathrsfs}
\usepackage{graphicx}
\usepackage{xcolor}

\usepackage[
colorlinks=true,
breaklinks=true,
citecolor=blue,
linkcolor=blue,
urlcolor=blue
]{hyperref}

\newcommand{\rmd}{\mathrm{d}}
\newcommand{\rmi}{\mathrm{i}}
\newcommand{\rme}{\mathrm{e}}

\newcommand{\rhoq}{\rho_{\mathrm{qubit}}}
\newcommand{\Psiq}{\Psi_{\mathrm{qubit}}}

\begin{document}

	\title{
	Fer expansion converges almost twice as far as Magnus for qubits
	}

	\author{Lorenzo Bagnasacco}
	\email{lorenzo.bagnasacco@sns.it}
	\affiliation{NEST and Scuola Normale Superiore, Piazza dei Cavalieri 7, I-56126 Pisa, Italy}

	\author{Vittorio Giovannetti}
	\affiliation{NEST, Scuola Normale Superiore and Istituto Nanoscienze-CNR, Piazza dei Cavalieri 7, I-56126 Pisa, Italy}

	\begin{abstract}
		The Fer expansion writes the propagator of a time-dependent Hamiltonian as a product of unitary exponentials whose error decreases doubly exponentially with the number of factors, while the Magnus expansion improves by one order at a time. Its convergence, however, was guaranteed only below a radius of about $2.6058$, smaller than the Magnus radius $\pi$. We establish a guaranteed convergence threshold of about $6.2106$, or $1.98\,\pi$, for a qubit. This is almost twice the optimal universal Magnus threshold $\pi$; the bound on a single Fer step is sharp, while the exact convergence radius remains open.
	\end{abstract}

	\maketitle

	\section{Introduction}
	\label{sec:intro}

	Time-dependent Hamiltonians are of fundamental relevance in quantum physics. A driven qubit is the basic model of magnetic resonance and of quantum computing~\cite{bloch,feynman,Vandersypen2005}, and simulating time-dependent Hamiltonian dynamics is a central task for quantum algorithms~\cite{kieferova2019,berry2020,casares2024,fang2025}. More generally, evolution with respect to parameters other than time also arises in a variety of quantum settings~\cite{spinscattering,oreshkov2009,sjoqvist2012}. For a time-dependent Hamiltonian $H(t)$ this dynamics is given formally by the time-ordered exponential \mbox{$U(t)=\mathcal{T}\exp[-\rmi\int_0^t\rmd s\,H(s)]$}, and expanding it in powers of $H$ gives the Dyson series~\cite{dyson}. The solution is only formal, since every truncation of the series is in general not unitary. How to reorganize it so that unitarity is preserved at every order is an old question, with two standard answers.

	\begin{figure}[t]
		\centering
		\includegraphics[width=\linewidth]{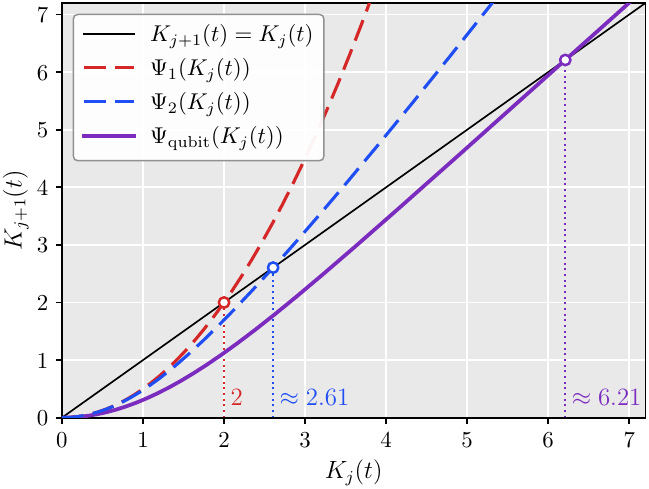}
		\caption{One step of the Fer recursion as a map $K_j(t)\mapsto K_{j+1}(t)$. The classical map $\Psi_1(x)$ (red, dashed) and the finite-angle map $\Psi_2(x)$ (blue, dashed) of Eq.~\eqref{eq:Krec} hold for Hermitian Hamiltonians of any dimension; the map $\Psiq(x)$ of Eq.~\eqref{eq:Phi} (violet) is the exact worst case of one step for a qubit. The guaranteed convergence thresholds are the positive fixed points, where the maps cross the diagonal (open circles).}
		\label{fig:maps}
	\end{figure}

	The first answer collects the whole Dyson series into one exponential, $U(t)=\rme^{\Omega(t)}$, where the anti-Hermitian operator $\Omega(t)$ is a series of nested commutators of $H$ at different times. This is the Magnus expansion~\cite{magnus,klarsfeld1989,blanes2009}. Each truncation is unitary by construction, but convergence of the series is not guaranteed for arbitrary times. Convergence is guaranteed when $\int_0^t\rmd s\,\|H(s)\|<\pi$~\cite{moan2006,casas2007}, where $\|\cdot\|$ denotes the operator norm, and $\pi$ is the best possible constant even for a qubit~\cite{lakos2019}.

	The second answer leaves the propagator as an ordered product of unitary exponentials. Examples include the Trotter--Suzuki formulas, built from compositions of exponentials and widely used in digital quantum simulation~\cite{suzuki,lloyd1996,childs2021}, and the Zassenhaus expansion~\cite{casas-zassenhaus} and its continuous version due to Wilcox~\cite{wilcox,arnal2021}. In the latter expansions, the $j$-th exponent is of order $j$ in the Hamiltonian, and successive factors add one order of accuracy at a time.

	The Fer expansion~\cite{fer,blanes1998} lies between these two answers. As in the Magnus expansion, only commutators of $H$ appear; as in the product formulas, the propagator is an ordered product. It is not, however, an expansion in powers of $H$. It extracts one exponential at a time: the exponent is the time integral of an effective Hamiltonian, and what remains obeys a genuine Schr\"odinger equation, with a new effective Hamiltonian built from the previous one. The new Hamiltonian is of second order in the previous one, so that each new factor roughly doubles the order in time to which the product reproduces the propagator, instead of adding one~\cite{iserles1984,iserles2000}. The expansion was rediscovered in numerical analysis by Iserles~\cite{iserles1984} and later recognized as a Lie-group method~\cite{iserles2000,zanna1999}. It is used to build effective Hamiltonians in solid-state nuclear magnetic resonance~\cite{madhu2006,mananga2016}, in the same spirit as average-Hamiltonian theory~\cite{haeberlen1968,mananga2011} and dynamical decoupling~\cite{viola1999}, and to construct geometric integrators~\cite{casas1996,zanna1999,zanna2001}.

	This fast convergence, however, is guaranteed only within a finite region. For a Hermitian Hamiltonian, the product was proved to converge when $\int_0^t\rmd s\,\|H(s)\|<2$~\cite{blanes1998}, and this value was recently raised to about $2.6058$~\cite{bagnasacco2026fer}. Both values are still smaller than the Magnus radius $\pi$. They come from the same argument, which follows a single number, the time integral $K_j(t)$ of the norm of the $j$-th effective Hamiltonian $H_j(t)$, and bounds $K_{j+1}(t)$ in terms of $K_j(t)$ by combining two inequalities. The first bounds the norm of $H_{j+1}(t)$ by the norm of $H_j(t)$ times a factor that increases with the norm of the time integral of $H_j(t)$. The second bounds the norm of this time integral by $K_j(t)$. The first inequality rests on an estimate for a single unitary conjugation that cannot be improved~\cite{stampfli1970}. For a qubit, however, the two inequalities are never tight at the same time. The norm of the time integral of $H_j(t)$ grows as fast as $K_j(t)$ only when $H_j(t)$ is proportional to it, and then $H_j(t)$ commutes with its time integral and gives no contribution to $H_{j+1}(t)$. A better radius thus requires keeping track of the correlation between an effective Hamiltonian and its own time integral.

	In this work we take this correlation into account exactly for a qubit, and show that the guaranteed threshold becomes more than twice as large as the general value of about $2.6058$. For a qubit, the effective Hamiltonians and their time integrals can be written as vectors of $\mathbb{R}^3$, and each step of the recursion is an explicit map of these vectors~\cite{bagnasacco2026su2}. The size of $H_{j+1}(t)$ turns out to be the size of the component of $H_j(t)$ orthogonal to its time integral, multiplied by a weight that depends only on the norm of the time integral. The time integral draws a curve of length $K_j(t)$ that starts at the origin, and the worst case of one step becomes a maximization over the shape of this curve, which we solve exactly. It gives the sufficient condition for convergence
	\begin{equation}
		\int_0^t\!\rmd s\;\frac{\lambda_{\max}(s)-\lambda_{\min}(s)}{2} < \rhoq \approx 6.2106
		\label{eq:intro}
	\end{equation}
	for every qubit Hamiltonian, with $\lambda_{\max}(s)$ and $\lambda_{\min}(s)$ its two eigenvalues. For a traceless Hamiltonian the integrand is $\|H(s)\|$. Notice that the guaranteed threshold $\rhoq$ is about $1.98\,\pi$, almost twice the optimal universal Magnus threshold $\pi$. The one-step bound is sharp, but this does not establish that $\rhoq$ is the exact convergence radius of the Fer expansion.
	
	The paper is organized as follows. Section~\ref{sec:fer} recalls the Fer recursion and the argument that turns a one-step bound into a sufficient condition for convergence. Section~\ref{sec:qubit} writes one step for a qubit and gives its exact size. Section~\ref{sec:worst} finds the worst case of one step: it writes the size of one step in terms of the radial speed of the curve (Sec.~\ref{sec:exactstep}), reduces the worst case to a problem on curves (Sec.~\ref{sec:curves}), and solves it (Sec.~\ref{sec:exact}): the Euler--Lagrange equation gives a candidate profile, and an inequality proves that it gives the worst case. Section~\ref{sec:attain} shows a qubit Hamiltonian that reaches it. Section~\ref{sec:radius} derives the corresponding sufficient condition for convergence and an error bound, and Sec.~\ref{sec:conclusions} concludes.

	\section{The Fer expansion}
	\label{sec:fer}

	We briefly recall the construction of the Fer expansion. Given a time-dependent Hamiltonian $H(t)=H^{\dagger}(t)$, the time-evolution operator $U(t)$ satisfies the Schr\"odinger equation ($\hbar=1$)
	\begin{equation}
		\frac{\rmd}{\rmd t}\,U(t) = -\rmi H(t)\,U(t)\;, \qquad U(0)=\openone\;.
		\label{eq:schrodinger}
	\end{equation}
	The Fer expansion~\cite{fer,blanes1998} writes it as an ordered product of exponentials,
	\begin{equation}
		U(t) = \rme^{F_1(t)}\rme^{F_2(t)}\cdots\rme^{F_j(t)}\,U_{j+1}(t)\;,
		\label{eq:product}
	\end{equation}
	where the remainder $U_{j+1}(t)$ obeys a Schr\"odinger equation of the same form, with the effective Hamiltonian
	\begin{equation}
		\begin{split}
			H_{j+1}(t) = \;&\rme^{-F_j(t)}\,H_j(t)\,\rme^{F_j(t)}\\
			&- \int_0^1\!\rmd\lambda\;\rme^{-\lambda F_j(t)}\,H_j(t)\,\rme^{\lambda F_j(t)}\;,
		\end{split}
		\label{eq:fer-recursion}
	\end{equation}
	where
	\begin{equation}
		F_j(t) = -\rmi\int_0^t\!\rmd s\;H_j(s)
		\label{eq:Fdef}
	\end{equation}
	and $H_1(t)\equiv H(t)$.

	The size of any Hamiltonian is measured by
	\begin{equation}
		K_j(t) \equiv \int_0^t\!\rmd s\;\big\|H_j(s)\big\|\;,
		\label{eq:Kdef}
	\end{equation}
	and the truncation error is bounded by the size of the first Hamiltonian that has been left out~\cite{blanes1998},
	\begin{equation}
		\Big\|U(t)-\rme^{F_1(t)}\cdots\rme^{F_N(t)}\Big\| \le K_{N+1}(t)\;.
		\label{eq:error}
	\end{equation}
	The expansion converges if $K_j(t)$ goes to zero as the number of factors grows.

	For Hermitian Hamiltonians of any dimension, Refs.~\cite{blanes1998} and~\cite{bagnasacco2026fer} prove two one-step bounds of the form
	\begin{equation}
		\big\|H_{j+1}(t)\big\| \le \psi_i\big(\|F_j(t)\|\big)\,\big\|H_j(t)\big\|\;,
		\label{eq:psi}
	\end{equation}
	with the classical coefficient $\psi_1(x)=x$ of Ref.~\cite{blanes1998} and the finite-angle coefficient of Ref.~\cite{bagnasacco2026fer},
	\begin{equation}
		\psi_2(x) =
		\begin{cases}
			\dfrac{2\,\big(1-\cos(x)\big)}{x}\;, & 0\le x\le\pi/2\;,\\[2ex]
			2-\dfrac{\pi-2}{x}\;, & x\ge\pi/2\;.
		\end{cases}
		\label{eq:psidef}
	\end{equation}
	Since $\|F_j(t)\|\le K_j(t)$ and $\psi_i(x)$ is increasing, integrating Eq.~\eqref{eq:psi} over time gives
	\begin{equation}
		K_{j+1}(t) \le \Psi_i\big(K_j(t)\big)\;, \qquad \Psi_i(x)\equiv\int_0^x\!\rmd u\;\psi_i(u)\;.
		\label{eq:Krec}
	\end{equation}
	Each map satisfies $\Psi_i(x)<x$ below its first positive fixed point $\rho$, which is the guaranteed convergence threshold of Refs.~\cite{blanes1998} and~\cite{bagnasacco2026fer}: $\rho=2$ for $\Psi_1(x)$ and $\rho$ about $2.6058$ for $\Psi_2(x)$. When $K_1(t)<\rho$ the sequence $K_j(t)$ decreases, and its limit can only be zero: the product~\eqref{eq:product} converges as the number of factors goes to infinity, with an error after $N$ factors not larger than the $N$-fold iterate $\Psi_i^{\circ N}\big(K_1(t)\big)$.

	\section{One step for a qubit}
	\label{sec:qubit}

	From now on we consider a qubit. In this case every effective Hamiltonian $H_j(t)$ and every exponent $F_j(t)$ is fixed by a vector of $\mathbb{R}^3$, and one step of the recursion~\eqref{eq:fer-recursion}--\eqref{eq:Fdef} becomes an explicit map between vectors~\cite{bagnasacco2026su2}. We first write the Hamiltonians as vectors (Sec.~\ref{sec:vectors}), then derive the map (Sec.~\ref{sec:map}), and finally read from it the exact size of one step (Sec.~\ref{sec:size}).

	\subsection{Hamiltonians as vectors}
	\label{sec:vectors}

	We consider a qubit Hamiltonian $H(t)=\mathbf{h}(t)\cdot\mathbf{T}$, where $\mathbf{h}(t)\in\mathbb{R}^3$ and $\mathbf{T}=(\mathrm{T}_1,\mathrm{T}_2,\mathrm{T}_3)$ is the vector of Pauli matrices, with $[\mathrm{T}_a,\mathrm{T}_b]=2\rmi\,\varepsilon_{abc}\,\mathrm{T}_c$~\cite{sakurai}. As shown in Ref.~\cite{bagnasacco2026su2}, the Fer recursion never leaves this algebra, and every effective Hamiltonian and every exponent can be written as
	\begin{equation}
		H_j(t) = \mathbf{h}_j(t)\cdot\mathbf{T}\;, \qquad
		F_j(t) = -\rmi\,\mathbf{f}_j(t)\cdot\mathbf{T}\;,
		\label{eq:vectors}
	\end{equation}
	with $\mathbf{h}_1(t)\equiv\mathbf{h}(t)$ and $\mathbf{f}_j(t)=\int_0^t\rmd s\,\mathbf{h}_j(s)$. Since $(\mathbf{x}\cdot\mathbf{T})^2=\|\mathbf{x}\|^2\openone$ for every $\mathbf{x}\in\mathbb{R}^3$, the operator norm of $\mathbf{x}\cdot\mathbf{T}$ is the Euclidean length $\|\mathbf{x}\|$ of $\mathbf{x}$. We write $\mathrm{h}_j(t)\equiv\|\mathbf{h}_j(t)\|$ and $\mathrm{f}_j(t)\equiv\|\mathbf{f}_j(t)\|$ for the lengths of the two vectors, so that Eq.~\eqref{eq:vectors} gives
	\begin{equation}
		\|H_j(t)\| = \mathrm{h}_j(t)\;, \qquad \|F_j(t)\| = \mathrm{f}_j(t)\;.
		\label{eq:norms}
	\end{equation}
	The size~\eqref{eq:Kdef} of the effective Hamiltonian is then the time integral of the length of $\mathbf{h}_j(t)$,
	\begin{equation}
		K_j(t) = \int_0^t\!\rmd s\;\mathrm{h}_j(s)\;.
		\label{eq:Kvec}
	\end{equation}
	This gives a geometric picture that we use throughout. As time goes on, the vector $\mathbf{f}_j(t)$ draws in $\mathbb{R}^3$ a curve that starts at the origin, since $\mathbf{f}_j(0)=\mathbf{0}$, and whose velocity is $\dot{\mathbf{f}}_j(t)=\mathbf{h}_j(t)$. Its distance from the origin is $\mathrm{f}_j(t)=\|F_j(t)\|$, and by Eq.~\eqref{eq:Kvec} its length up to the time $t$ is $K_j(t)$. Since a curve is never shorter than the segment that joins its end points, $\mathrm{f}_j(t)\le K_j(t)$, which is the inequality $\|F_j(t)\|\le K_j(t)$ used in Sec.~\ref{sec:fer}.
	
	\begin{figure}[!t]
		\centering
		\includegraphics[width=\linewidth]{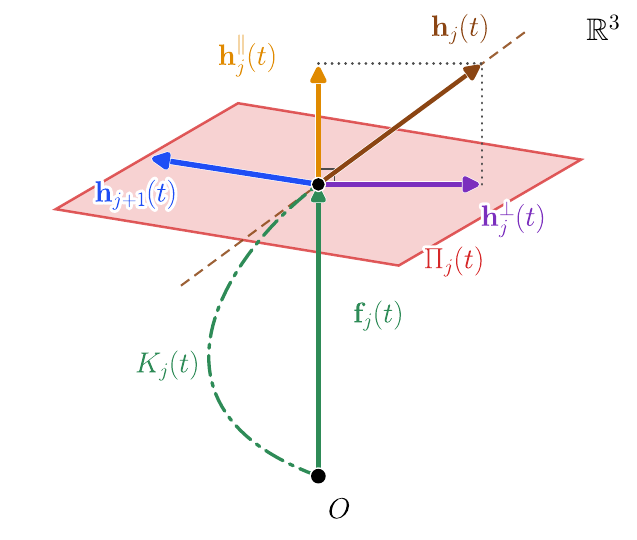}
		\caption{The curve $\mathbf{f}_j(s)$, $0\le s\le t$ (green, dash-dotted), has length $K_j(t)$ and ends at the distance $\mathrm{f}_j(t)$ from the origin $O$. At its end point the velocity $\mathbf{h}_j(t)$ (brown) is tangent to the curve, along the brown dashed line, and it splits into the radial part $\mathbf{h}^{\parallel}_j(t)$ (orange), along $\mathbf{f}_j(t)$, and the orthogonal part $\mathbf{h}^{\perp}_j(t)$ (violet), in the plane $\Pi_j(t)$ orthogonal to $\mathbf{f}_j(t)$ (red). One step of the Fer recursion erases the radial part: the new vector $\mathbf{h}_{j+1}(t)$ (blue) lies in $\Pi_j(t)$, and it is the orthogonal part rotated around $\mathbf{f}_j(t)$ and multiplied by $\eta\big(\mathrm{f}_j(t)\big)$.}
		\label{fig:geomcurve}
	\end{figure}

	\subsection{The Fer map}
	\label{sec:map}

	One step of the Fer recursion maps the vector $\mathbf{h}_j(t)$ into the vector $\mathbf{h}_{j+1}(t)$, and in this subsection we write this map explicitly. For a qubit the conjugations in Eq.~\eqref{eq:fer-recursion} are rotations of $\mathbb{R}^3$, given explicitly by the Rodrigues formula~\cite{sakurai}. Inserting it into Eq.~\eqref{eq:fer-recursion}, one shows~\cite{bagnasacco2026su2} that the new vector $\mathbf{h}_{j+1}(t)$ is a combination of two auxiliary vectors. The first is the cross product of $\mathbf{f}_j(t)$ with $\mathbf{h}_j(t)$, and the second is the cross product of the first with $\mathbf{f}_j(t)$,
	\begin{align}
		\mathbf{p}_{j+1}(t)&\equiv\mathbf{f}_j(t)\times\mathbf{h}_j(t)\;, \label{eq:pq}\\
		\mathbf{q}_{j+1}(t)&\equiv\mathbf{p}_{j+1}(t)\times\mathbf{f}_j(t)\;. \label{eq:q}
	\end{align}
	In terms of these vectors the map reads
	\begin{equation}
		\mathbf{h}_{j+1}(t) = \alpha\big(\mathrm{f}_j(t)\big)\,\mathbf{p}_{j+1}(t) + \beta\big(\mathrm{f}_j(t)\big)\,\mathbf{q}_{j+1}(t)\;,
		\label{eq:map}
	\end{equation}
	with the weights
	\begin{align}
		\alpha(x) &= \frac{1-\cos(2x)}{2x^2} - \frac{\sin(2x)}{x}\;, \label{eq:alpha}\\
		\beta(x) &= \frac{\cos(2x)}{x^2} - \frac{\sin(2x)}{2x^3}\;. \label{eq:beta}
	\end{align}
	Figure~\ref{fig:geomcurve} shows the map~\eqref{eq:map} in the geometric picture of Sec.~\ref{sec:vectors}. Up to the time $t$ the curve $\mathbf{f}_j(s)$, $0\le s\le t$, has length $K_j(t)$ and ends at the distance $\mathrm{f}_j(t)$ from the origin, where its velocity $\mathbf{h}_j(t)$ is tangent to it. We split the velocity into a radial part along $\mathbf{f}_j(t)$, $\mathbf{h}^{\parallel}_j(t)\equiv[\mathbf{f}_j(t)\cdot\mathbf{h}_j(t)]\,\mathbf{f}_j(t)/\mathrm{f}^2_j(t)$, and an orthogonal part $\mathbf{h}^{\perp}_j(t)\equiv\mathbf{h}_j(t)-\mathbf{h}^{\parallel}_j(t)$, which lies in the plane $\Pi_j(t)$ orthogonal to $\mathbf{f}_j(t)$. The radial part drops out of the cross product~\eqref{eq:pq}, so that $\mathbf{p}_{j+1}(t)=\mathbf{f}_j(t)\times\mathbf{h}^{\perp}_j(t)$ is the orthogonal part turned by a right angle around $\mathbf{f}_j(t)$ and multiplied by $\mathrm{f}_j(t)$. Expanding the double cross product~\eqref{eq:q} gives instead $\mathbf{q}_{j+1}(t)=\mathrm{f}^2_j(t)\,\mathbf{h}_j(t)-[\mathbf{f}_j(t)\cdot\mathbf{h}_j(t)]\,\mathbf{f}_j(t)=\mathrm{f}^2_j(t)\,\mathbf{h}^{\perp}_j(t)$. Both auxiliary vectors lie in $\Pi_j(t)$, and so does their combination $\mathbf{h}_{j+1}(t)$~\cite{bagnasacco2026su2}. One step of the recursion thus erases the radial part of $\mathbf{h}_j(t)$, while it rotates the orthogonal part around $\mathbf{f}_j(t)$ and changes its length.

	\subsection{The size of one step}
	\label{sec:size}

	The two terms of Eq.~\eqref{eq:map} are orthogonal, since $\mathbf{q}_{j+1}(t)$ is parallel to the orthogonal part $\mathbf{h}^{\perp}_j(t)$, while $\mathbf{p}_{j+1}(t)$ is $\mathbf{h}^{\perp}_j(t)$ turned by a right angle. Their lengths are $\|\mathbf{p}_{j+1}(t)\|=\mathrm{f}_j(t)\,\|\mathbf{h}^{\perp}_j(t)\|$ and $\|\mathbf{q}_{j+1}(t)\|=\mathrm{f}^2_j(t)\,\|\mathbf{h}^{\perp}_j(t)\|$, so that by Pythagoras the length of $\mathbf{h}_{j+1}(t)$ is
	\begin{equation}
		\mathrm{h}_{j+1}(t) = \eta\big(\mathrm{f}_j(t)\big)\,\|\mathbf{h}^{\perp}_j(t)\|\;,
		\label{eq:size}
	\end{equation}
	with the weight
	\begin{equation}
		\begin{split}
			\eta^2(x) &\equiv x^2\big[\alpha^2(x)+x^2\beta^2(x)\big]\\
			&= 1-\frac{\sin(2x)}{x}+\frac{\sin^2(x)}{x^2}\;.
		\end{split}
		\label{eq:eta}
	\end{equation}
	This is exactly what one step of the recursion does to the size of a qubit Hamiltonian. In Fig.~\ref{fig:geomcurve}, the new vector $\mathbf{h}_{j+1}(t)$ is the orthogonal part $\mathbf{h}^{\perp}_j(t)$ rotated around $\mathbf{f}_j(t)$ and multiplied by the weight $\eta\big(\mathrm{f}_j(t)\big)$, which depends only on the distance $\mathrm{f}_j(t)$ of the curve from the origin.
	The weight $\eta(x)$, shown in Fig.~\ref{fig:etaL}, increases from $\eta(0)=0$ up to its maximum, reached at
	\begin{equation}
		x^\star \approx 2.04\;, \qquad \eta^\star\equiv\eta(x^\star) \approx 1.26\;.
		\label{eq:xstar}
	\end{equation}
	For larger $x$ the weight oscillates around $1$ and approaches $1$ as $x\to\infty$.
	

	\section{The worst case of one step}
	\label{sec:worst}
	
	As in Refs.~\cite{blanes1998,bagnasacco2026fer}, our goal is to express the size $K_{j+1}(t)$ of the next effective Hamiltonian in terms of the size $K_j(t)$ of the current one, and then to look for the fixed points of the resulting map, which yield a sufficient condition for convergence.

	\subsection{An exact formula for one step}
	\label{sec:exactstep}
	
	For a qubit, Eq.~\eqref{eq:size} already gives the size of one step exactly, but in terms of the component of $\mathbf{h}_j(t)$ orthogonal to $\mathbf{f}_j(t)$. In this subsection we rewrite it in terms of the speed $\mathrm{h}_j(t)$ of the curve $\mathbf{f}_j(t)$, whose time integral is $K_j(t)$, and of its distance $\mathrm{f}_j(t)$ from the origin.

	We differentiate $\mathrm{f}^2_j(t)=\mathbf{f}_j(t)\cdot\mathbf{f}_j(t)$ with $\dot{\mathbf{f}}_j(t)=\mathbf{h}_j(t)$, and obtain the speed at which the curve $\mathbf{f}_j(t)$ moves away from the origin, the radial speed
	\begin{equation}
		\dot{\mathrm{f}}_j(t) = \frac{\mathbf{f}_j(t)\cdot\mathbf{h}_j(t)}{\mathrm{f}_j(t)}\;,
		\label{eq:radial}
	\end{equation}
	which is the component of the velocity $\mathbf{h}_j(t)$ along the unit vector $\mathbf{f}_j(t)/\mathrm{f}_j(t)$. A component of a vector is never longer than the vector itself, $|\mathbf{f}_j(t)\cdot\mathbf{h}_j(t)|\le\mathrm{f}_j(t)\,\mathrm{h}_j(t)$ by the Cauchy--Schwarz inequality, and therefore $|\dot{\mathrm{f}}_j(t)|\le\mathrm{h}_j(t)$: the distance from the origin cannot change faster than the curve moves, and it changes at the speed $\mathrm{h}_j(t)$ only when $\mathbf{h}_j(t)$ is parallel to $\mathbf{f}_j(t)$. The radial part $\mathbf{h}^{\parallel}_j(t)$ of Sec.~\ref{sec:map} has length $|\dot{\mathrm{f}}_j(t)|$, so that by Pythagoras the orthogonal part has length $\|\mathbf{h}^{\perp}_j(t)\|=\Big[\mathrm{h}^2_j(t)-\big(\dot{\mathrm{f}}_j(t)\big)^2\Big]^{1/2}$. Together with Eqs.~\eqref{eq:norms} and~\eqref{eq:size}, this yields the exact size of one step,
	\begin{equation}
		\mathrm{h}_{j+1}(t) = \eta\big(\mathrm{f}_j(t)\big)\,\sqrt{\mathrm{h}^2_j(t)-\big(\dot{\mathrm{f}}_j(t)\big)^2}\;,
		\label{eq:exact}
	\end{equation}
	and, integrating over time and using $\mathrm{h}_j(s)=\dot K_j(s)$ from Eq.~\eqref{eq:Kvec},
	\begin{equation}
		K_{j+1}(t) = \int_0^t\!\rmd s\;\eta\big(\mathrm{f}_j(s)\big)\,\sqrt{\big(\dot K_j(s)\big)^2-\big(\dot{\mathrm{f}}_j(s)\big)^2}\;.
		\label{eq:exactK}
	\end{equation}
	At the times when $\mathbf{f}_j(t)=\mathbf{0}$ the radial speed is not defined, but both sides of Eq.~\eqref{eq:exact} vanish, since $\eta(0)=0$.

	\begin{figure}[!t]
		\centering
		\includegraphics[width=\linewidth]{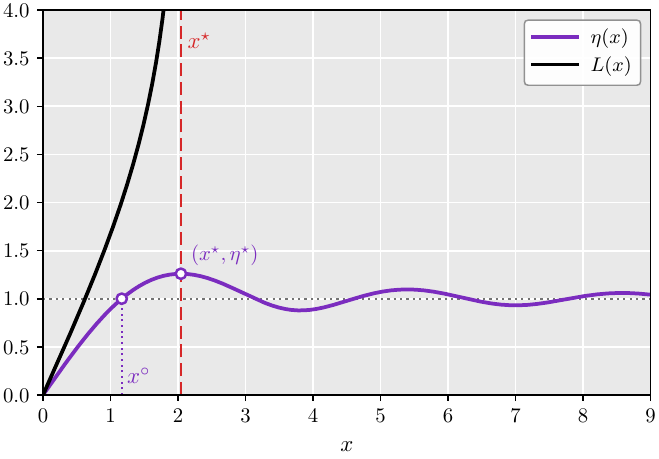}
		\caption{The two functions that determine the optimal radial profile in Sec.~\ref{sec:candidate}. The weight $\eta(x)$ of Eq.~\eqref{eq:eta} (violet) reaches $1$ for the first time at $x^\circ$ (open circle), increases up to its maximum $\eta^\star$ at $x^\star$ (red, dashed), then oscillates around $1$ and tends to $1$ for large $x$ (dotted line). The length $L(x)$ of Eq.~\eqref{eq:Ldef} (black) is defined for $0\le x<x^\star$: it grows from $0$, as $\pi x/2$ for small $x$, and diverges at $x=x^\star$.}
		\label{fig:etaL}
	\end{figure}

	\subsection{The worst case as a problem on curves}
	\label{sec:curves}

	In Eq.~\eqref{eq:exactK} we now take as integration variable the length $\sigma$ traveled by the curve up to the time $s$,
	\begin{equation}
		\sigma \equiv K_j(s) = \int_{0}^{s}\!\rmd s'\,\|\dot{\mathbf{f}}_j(s')\|\;,
		\label{eq:sigmadef}
	\end{equation}
	and describe the curve by its radial profile $r(\sigma)$, the distance from the origin after the length $\sigma$ has been traveled,
	\begin{equation}
		r(\sigma) \equiv \mathrm{f}_j(s)\;.
		\label{eq:sigma}
	\end{equation}
	Changing the integration variable of Eq.~\eqref{eq:exactK} from $s$ to $\sigma$, with $\rmd\sigma=\dot K_j(s)\,\rmd s$ and $\dot{\mathrm{f}}_j(s)=\big(\rmd r(\sigma)/\rmd\sigma\big)\,\dot K_j(s)$, gives
	\begin{equation}
		K_{j+1}(t) = \int_0^{K_j(t)}\!\rmd\sigma\;\eta\big(r(\sigma)\big)\sqrt{1-\Big(\frac{\rmd r(\sigma)}{\rmd\sigma}\Big)^2}\;,
		\label{eq:compact}
	\end{equation}
	where the square root is real: by Eq.~\eqref{eq:radial} the radial speed is at most the speed, $|\rmd\mathrm{f}_j(s)/\rmd s|\le\mathrm{h}_j(s)=\dot K_j(s)$, and comparing with $\rmd\mathrm{f}_j(s)/\rmd s=\big(\rmd r(\sigma)/\rmd\sigma\big)\,\dot K_j(s)$ one has $|\rmd r(\sigma)/\rmd\sigma|\le1$, that is, along the curve the distance from the origin cannot change by more than the length traveled. Since it is a distance that starts from $\mathrm{f}_j(0)=0$, every radial profile therefore satisfies the constraints
	\begin{equation}\label{eq:constraints_r}
		r(0)=0\;,\quad r(\sigma)\ge0\;,\quad \Big|\frac{\rmd r(\sigma)}{\rmd\sigma}\Big|\le1\;.
	\end{equation}
	For a curve of length $\Sigma$ we call admissible the radial profiles that satisfy these constraints, and denote their set by
	\begin{equation}
		\begin{split}
			\mathbb{X}_\Sigma\equiv\big\{&r:[0,\Sigma]\to\mathbb{R}\;:\;r(0)=0\;,\;r(\sigma)\ge0\;,\\
			&|r(\sigma)-r(\tau)|\le|\sigma-\tau|\;\;\forall\,\sigma,\tau\big\}\;.
		\end{split}
		\label{eq:XSigma}
	\end{equation}
	The last condition is the last constraint of Eq.~\eqref{eq:constraints_r} written without derivatives: it also admits profiles with corners, which have a derivative at almost every $\sigma$. Equation~\eqref{eq:compact} shows that $K_{j+1}(t)$ depends on $K_j(t)$ only through the upper limit of integration and on the Hamiltonian only through the radial profile $r(\sigma)$ of the curve.


	\subsection{Solving the problem on curves}
	\label{sec:exact}

	Equation~\eqref{eq:compact} is an equality: for a given Hamiltonian, $K_{j+1}(t)$ is exactly the integral on its right-hand side. Hamiltonians with the same $K_j(t)$ have in general different radial profiles, and knowing $K_j(t)$ alone we do not know which one applies. As in Sec.~\ref{sec:fer}, however, we look for a bound on $K_{j+1}(t)$ that involves $K_j(t)$ only. Such a bound must hold for every radial profile allowed by the constraints~\eqref{eq:constraints_r}, that is, in the worst case, and we want it to be as small as possible, since a smaller bound gives a larger convergence radius. We therefore fix the length $K_j(t)$ to an arbitrary value $\Sigma$ and look for the largest value of the functional
	\begin{equation}
		\mathscr{F}[r]\equiv\int_0^{\Sigma}\!\rmd\sigma\;\mathscr{L}\big(\sigma,r(\sigma),\dot r(\sigma)\big)
		\label{eq:J}
	\end{equation}
	over the admissible profiles $r\in\mathbb{X}_\Sigma$ of Eq.~\eqref{eq:XSigma}, with the final distance $r(\Sigma)$ free. Its Lagrangian is
	\begin{equation}
		\mathscr{L}\big(\sigma,r(\sigma),\dot r(\sigma)\big)\equiv\eta\big(r(\sigma)\big)\sqrt{1-\big(\dot r(\sigma)\big)^2}\;,
		\label{eq:Lag}
	\end{equation}
	where $\dot r(\sigma)\equiv\rmd r(\sigma)/\rmd\sigma$ is the radial speed, the derivative of the profile with respect to the length traveled. In this subsection the dot is a derivative with respect to $\sigma$ and no longer with respect to time, while the prime denotes the derivative of a function with respect to its argument, as in $\eta'(x)\equiv\rmd\eta(x)/\rmd x$. The integrand contains two competing factors. The weight $\eta\big(r(\sigma)\big)$ is large only when the curve is far from the origin, at distances close to $x^\star$, while the square root is large only when the distance $r(\sigma)$ changes slowly.

	We find the largest value in two parts, as is usual in the calculus of variations~\cite{gelfand1963,buttazzo1998,gobbino2016}. In the first part (Sec.~\ref{sec:candidate}) we assume that the largest value exists and is reached by a regular profile, and the Euler--Lagrange equation gives a candidate profile $\bar r(\sigma)$. These assumptions only serve to find the candidate, and nothing is proved there. In the second part (Sec.~\ref{sec:proof}) we prove, with an inequality that holds for every admissible profile, that the candidate gives the largest value. This also shows that the largest value exists.

	\subsubsection{A candidate from the Euler--Lagrange equation}
	\label{sec:candidate}

	Suppose that the largest value of $\mathscr{F}[r]$ is reached by a profile $r\in\mathbb{X}_\Sigma$. We compare it with the nearby profiles $r(\sigma)+\varepsilon w(\sigma)$, where $\varepsilon$ is a small number of either sign and $w(\sigma)$ is a differentiable function with $w(0)=0$, so that the starting point does not move, and we suppose that these profiles are admissible. The final value $w(\Sigma)$ is free, as $r(\Sigma)$ is. Along this family, $\mathscr{F}$ becomes a function of $\varepsilon$ alone, $\varphi(\varepsilon)\equiv \mathscr{F}[r+\varepsilon w]$, which is largest at $\varepsilon=0$, so that its derivative vanishes there.

	We compute this derivative by differentiating under the integral at fixed $\sigma$ and setting $\varepsilon=0$, which gives
	\begin{multline}
		\frac{\rmd\varphi(\varepsilon)}{\rmd\varepsilon}\Big|_{\varepsilon=0}=-\int_0^{\Sigma}\!\rmd\sigma\;\frac{\eta\big(r(\sigma)\big)\,\dot r(\sigma)}{\sqrt{1-\big(\dot r(\sigma)\big)^2}}\,\dot w(\sigma)\\
		+\int_0^{\Sigma}\!\rmd\sigma\;\eta'\big(r(\sigma)\big)\sqrt{1-\big(\dot r(\sigma)\big)^2}\,w(\sigma)\;.
		\label{eq:dphi}
	\end{multline}
	In the first integral $\dot w(\sigma)$ is multiplied by $-P(\sigma)$, with
	\begin{equation}
		P(\sigma)\equiv\frac{\eta\big(r(\sigma)\big)\,\dot r(\sigma)}{\sqrt{1-\big(\dot r(\sigma)\big)^2}}\;.
		\label{eq:P}
	\end{equation}
	An integration by parts moves the derivative from $w(\sigma)$ to $P(\sigma)$. This requires $r(\sigma)$ to have two continuous derivatives, since $P(\sigma)$ contains $\dot r(\sigma)$. The boundary term at $\sigma=0$ vanishes, because $w(0)=0$, while the one at $\sigma=\Sigma$ remains:
	\begin{multline}
		\frac{\rmd\varphi(\varepsilon)}{\rmd\varepsilon}\Big|_{\varepsilon=0}=-P(\Sigma)\,w(\Sigma)+\int_0^{\Sigma}\!\rmd\sigma\;\dot P(\sigma)\,w(\sigma)\\
		+\int_0^{\Sigma}\!\rmd\sigma\;\eta'\big(r(\sigma)\big)\sqrt{1-\big(\dot r(\sigma)\big)^2}\,w(\sigma)\;.
		\label{eq:firstvar}
	\end{multline}

	This derivative vanishes for every $w(\sigma)$, and we use this fact in two steps. We first take $w(\Sigma)=0$. The boundary term disappears, and the two integrals together vanish for every $w(\sigma)$ that vanishes at both ends. By the fundamental lemma of the calculus of variations the sum of their integrands then vanishes at every $\sigma$,
	\begin{equation}
		\dot P(\sigma)=-\eta'\big(r(\sigma)\big)\sqrt{1-\big(\dot r(\sigma)\big)^2}\;.
		\label{eq:Puse}
	\end{equation}
	Computing $\dot P(\sigma)$ from the definition~\eqref{eq:P} with the product rule, Eq.~\eqref{eq:Puse} becomes the Euler--Lagrange equation
	\begin{equation}
		\eta\big(r(\sigma)\big)\,\ddot r(\sigma)=-\eta'\big(r(\sigma)\big)\,\big[1-\big(\dot r(\sigma)\big)^2\big]\;.
		\label{eq:ELE}
	\end{equation}
	We then take $w(\Sigma)\neq0$. The two integrals in Eq.~\eqref{eq:firstvar} cancel, only the boundary term is left, and $P(\Sigma)=0$. Excluding the case $r(\Sigma)=0$, in which the curve ends at the origin and the weight vanishes, this gives $\dot r(\Sigma)=0$: since the final distance is free, the profile reaches it with zero radial speed.

	The Lagrangian~\eqref{eq:Lag} does not depend on $\sigma$ explicitly, so that the Euler--Lagrange equation has a constant of motion, a fact known as the Beltrami identity~\cite{gelfand1963,gobbino2016}. With Eq.~\eqref{eq:P}, it is $E(\sigma)\equiv\mathscr{L}\big(\sigma,r(\sigma),\dot r(\sigma)\big)+\dot r(\sigma)P(\sigma)=\eta\big(r(\sigma)\big)\big[1-\big(\dot r(\sigma)\big)^2\big]^{-1/2}$. Computing it at $\sigma=\Sigma$, where $\dot r(\Sigma)=0$, we get
	\begin{equation}
		\frac{\eta\big(r(\sigma)\big)}{\sqrt{1-\big(\dot r(\sigma)\big)^2}}=\eta\big(r(\Sigma)\big)\;.
		\label{eq:endvalue}
	\end{equation}
	Solving for $\dot r(\sigma)$, with the positive root since the profile moves away from the origin, we find the radial profile
	\begin{equation}
		\dot r(\sigma)=\sqrt{1-\frac{\eta^2\big(r(\sigma)\big)}{\eta^2\big(r(\Sigma)\big)}}\;.
		\label{eq:spiral}
	\end{equation}
	At $\sigma=\Sigma$, Eq.~\eqref{eq:spiral} gives $\dot r(\Sigma)=0$, the condition found above for a free final distance. Separating the variables in Eq.~\eqref{eq:spiral} and integrating from $0$ to $\sigma$, we obtain the length that the profile needs to reach the distance $r(\sigma)$,
	\begin{equation}
		\sigma=\int_0^{r(\sigma)}\!\rmd u\;\frac{\eta\big(r(\Sigma)\big)}{\sqrt{\eta^2\big(r(\Sigma)\big)-\eta^2(u)}}\;.
		\label{eq:sigmar}
	\end{equation}
	At $\sigma=\Sigma$, Eq.~\eqref{eq:sigmar} becomes a condition on the final distance,
	\begin{align}
		L\big(r(\Sigma)\big)&=\Sigma\;,\label{eq:L}\\
		L(x)&\equiv\int_0^{x}\!\rmd u\;\frac{\eta(x)}{\sqrt{\eta^2(x)-\eta^2(u)}}\;.\label{eq:Ldef}
	\end{align}
	As shown in Fig.~\ref{fig:etaL}, $L(x)$ increases from zero and diverges as $x$ approaches $x^\star$. The condition $L\big(r(\Sigma)\big)=\Sigma$ therefore fixes a single final distance $r(\Sigma)$ in $(0,x^\star)$.


	In summary, the maximizing candidate profile $\bar r(\sigma)$ must belong to $\mathbb{X}_\Sigma$ and, moreover, satisfy
	\begin{equation}
		\dot{\bar r}(\sigma)=\sqrt{1-\frac{\eta^2\big(\bar r(\sigma)\big)}{\eta^2\big(\bar r(\Sigma)\big)}}\;,\qquad L\big(\bar r(\Sigma)\big)=\Sigma\;.
		\label{eq:candidate}
	\end{equation}
	The first equation contains the Euler--Lagrange equation~\eqref{eq:ELE}. At the end of the profile, the first and the second equation give, respectively,
	\begin{equation}
		\dot{\bar r}(\Sigma)=0\;,\qquad \bar r(\Sigma)\in(0,x^{\star})\;.
		\label{eq:candidateends}
	\end{equation}

	We have not justified any of the assumptions made in this part: the candidate still has to be shown to give the largest value. The second part does this without any of these assumptions, and uses the previous conditions only as a starting point.

	\begin{figure}[!t]
		\centering
		\includegraphics[width=\linewidth,trim=12bp 0 0 0,clip]{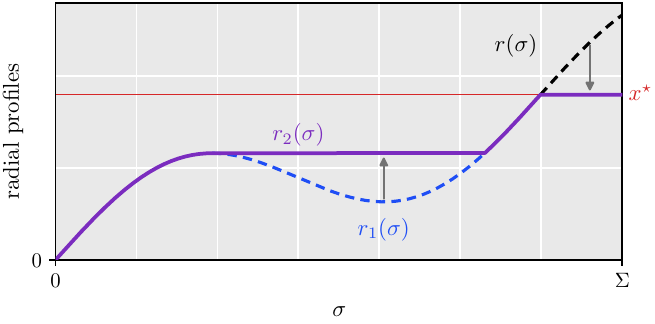}
		\caption{The first step of the proof in Sec.~\ref{sec:proof}, for an example radial profile. The profile $r(\sigma)$ differs from $r_1(\sigma)$ of Eq.~\eqref{eq:step1} only where it exceeds $x^\star$ (red): there it is the black dashed line, and $r_1(\sigma)=x^\star$. The profile $r_1(\sigma)$ differs from $r_2(\sigma)$ of Eq.~\eqref{eq:step1b} (violet) only where it lies below the largest value reached before: there it is the blue dashed line, and $r_2(\sigma)$ stays at that value. The arrows show the two modifications. On both the derivative vanishes and the integrand of Eq.~\eqref{eq:J} does not decrease, which gives Eq.~\eqref{eq:step1ineq}.}
		\label{fig:step1compact}
	\end{figure}

	\subsubsection{The candidate gives the largest value}
	\label{sec:proof}

	We now prove that $\mathscr{F}[r]\le \mathscr{F}[\bar r]$ for every admissible profile $r(\sigma)$. Nothing of the first part is used, except the candidate itself.

	\textit{First step: monotone profiles below $x^\star$.} The first of Eqs.~\eqref{eq:candidate} and the second of Eqs.~\eqref{eq:candidateends} suggest which profiles to consider. The final distance of the candidate lies below $x^\star$, and on $[0,x^\star]$ the weight $\eta(x)$ is strictly increasing, so that the argument of the square root in Eq.~\eqref{eq:candidate} is nonnegative only if the final distance is the largest one. We now obtain both properties for a generic admissible profile $r(\sigma)$: we first cut it at $x^\star$ and then replace it with its running maximum (Fig.~\ref{fig:step1compact}),
	\begin{align}
		r_1(\sigma)&\equiv\min\big\{r(\sigma),x^\star\big\}\;,\label{eq:step1}\\
		r_2(\sigma)&\equiv\max_{0\le\tau\le\sigma}r_1(\tau)\;.\label{eq:step1b}
	\end{align}
	Where $r_1(\sigma)$ differs from $r(\sigma)$, the profile $r_1(\sigma)$ stays at the distance $x^\star$: its derivative vanishes, and its weight $\eta(x^\star)$ is the largest value of $\eta(x)$. Where $r_2(\sigma)$ differs from $r_1(\sigma)$, the profile $r_2(\sigma)$ stays at the largest distance reached before: its derivative vanishes again, and its weight is not smaller, because $\eta(x)$ is increasing on $[0,x^\star]$. In both cases the square root takes its largest value $1$, so that at almost every $\sigma$ the integrand of Eq.~\eqref{eq:J} does not decrease, and
	\begin{equation}
		\mathscr{F}[r]\le \mathscr{F}[r_1]\le \mathscr{F}[r_2]\;.
		\label{eq:step1ineq}
	\end{equation}
	The profile $r_2(\sigma)$ is still admissible, it never decreases, and it never exceeds $x^\star$. We call $x_{\mathrm{f}}\equiv r_2(\Sigma)$, in $[0,x^\star]$, its final distance. It is the largest distance from the origin reached by the curve, cut at $x^\star$, and in general it differs from the final distance $r(\Sigma)$ of the original profile. Since $r_2(\sigma)$ never decreases, $0\le r_2(\sigma)\le x_{\mathrm{f}}$ for every $\sigma$. If $x_{\mathrm{f}}=0$, then $r_2(\sigma)$ vanishes and $\mathscr{F}[r]\le \mathscr{F}[r_2]=0\le \mathscr{F}[\bar r]$, so that from now on we take $x_{\mathrm{f}}>0$, and $\eta(x_{\mathrm{f}})>0$.

\begin{figure}[!t]
		\centering
		\includegraphics[width=\linewidth]{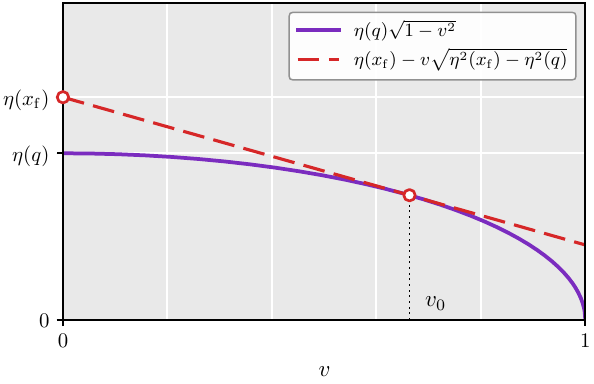}
		\caption{The second step of the proof in Sec.~\ref{sec:proof}. At fixed $q$, the Lagrangian $\mathscr{L}(\sigma,q,v)=\eta(q)\sqrt{1-v^2}$ is concave in $v$ (violet), and its graph lies below its tangent line $\ell(v)$ of Eq.~\eqref{eq:tangent} at the point $v_0$ of Eq.~\eqref{eq:v0} (red, dashed; open circle). The tangent line takes the value $\eta(x_{\mathrm{f}})$ at $v=0$ (open circle). At every $v$ the vertical distance between the two is the right-hand side of Eq.~\eqref{eq:cs} minus its left-hand side, which is therefore nonnegative.}
		\label{fig:step2}
	\end{figure}

	\textit{Second step: an inequality between numbers.} We consider the Lagrangian~\eqref{eq:Lag} at a position $q$ in $(0,x_{\mathrm{f}}]$ and a velocity $v$ in $[0,1]$, two numbers that will play the role of $r(\sigma)$ and $\dot r(\sigma)$, so that $\mathscr{L}(\sigma,q,v)=\eta(q)\sqrt{1-v^2}$. At fixed $q$ this is a concave function of $v$, so that its graph lies below each of its tangent lines (Fig.~\ref{fig:step2}). We take the tangent at the velocity that the candidate would have at the position $q$, that is
	\begin{equation}
		v_0\equiv\sqrt{1-\frac{\eta^2(q)}{\eta^2(x_{\mathrm{f}})}}\;.
		\label{eq:v0}
	\end{equation}
	The tangent line at $v_0$ is
	\begin{equation}
		\begin{split}
			\ell(v)&\equiv\frac{\partial\mathscr{L}(\sigma,q,v)}{\partial v}\Big|_{v=v_0}(v-v_0)+\mathscr{L}(\sigma,q,v_0)\\
			&=\eta(x_{\mathrm{f}})-v\sqrt{\eta^2(x_{\mathrm{f}})-\eta^2(q)}\;,
		\end{split}
		\label{eq:tangent}
	\end{equation}
	and by concavity $\mathscr{L}(\sigma,q,v)\le\ell(v)$ for every $v$ in $[0,1]$, with equality only for $v=v_0$. This is the inequality between numbers
	\begin{equation}
		\eta(q)\sqrt{1-v^2}+v\sqrt{\eta^2(x_{\mathrm{f}})-\eta^2(q)}\le\eta(x_{\mathrm{f}})\;.
		\label{eq:cs}
	\end{equation}
	For $q=0$ the weight vanishes and Eq.~\eqref{eq:cs} reads $v\,\eta(x_{\mathrm{f}})\le\eta(x_{\mathrm{f}})$, which holds for every $v$ in $[0,1]$.

	\textit{Third step: integration.} We set $q=r_2(\sigma)$ and $v=\dot r_2(\sigma)$ in Eq.~\eqref{eq:cs}, which is allowed since $0\le r_2(\sigma)\le x_{\mathrm{f}}$ and $0\le\dot r_2(\sigma)\le1$, and integrate from $\sigma=0$ to $\sigma=\Sigma$. The first term gives $\mathscr{F}[r_2]$, and the right-hand side gives $\eta(x_{\mathrm{f}})\Sigma$. Changing the integration variable to $u=r_2(\sigma)$, the second term becomes an integral from $0$ to $x_{\mathrm{f}}$, which depends only on $x_{\mathrm{f}}$. We write it as $I(x_{\mathrm{f}})$, with
	\begin{equation}
		I(x)\equiv\int_0^{x}\!\rmd u\;\sqrt{\eta^2(x)-\eta^2(u)}\;.
		\label{eq:Idef}
	\end{equation}
	Together with the first step,
	\begin{equation}
		\mathscr{F}[r]\le \mathscr{F}[r_2]\le\eta(x_{\mathrm{f}})\,\Sigma-I(x_{\mathrm{f}})\;.
		\label{eq:I}
	\end{equation}
	The two terms of this bound have a simple meaning. The first, $\eta(x_{\mathrm{f}})\Sigma$, is what one would obtain if the curve could spend its whole length at the distance $x_{\mathrm{f}}$. The second, $I(x_{\mathrm{f}})$, is the price paid for the part of the length spent moving away from the origin to reach that distance.

	\textit{Fourth step: the worst final distance.} The distance $x_{\mathrm{f}}$ depends on the profile, and we remove it by taking the largest value of $g(x)\equiv\eta(x)\Sigma-I(x)$ over $[0,x^\star]$. To find it we differentiate $g(x)$. In $I(x)$ the variable $x$ appears both in the upper limit and in the integrand. The upper limit gives the integrand at $u=x$, which vanishes, and the integrand gives
	\begin{equation}
		\frac{\rmd I(x)}{\rmd x}=\int_0^{x}\!\rmd u\;\frac{\eta(x)\,\eta'(x)}{\sqrt{\eta^2(x)-\eta^2(u)}}=\eta'(x)\,L(x)\;,
		\label{eq:dI}
	\end{equation}
	so that
	\begin{equation}
		\frac{\rmd g(x)}{\rmd x}=\eta'(x)\,\big[\Sigma-L(x)\big]\;.
		\label{eq:derivative}
	\end{equation}
	In the interval $(0,x^\star)$ we have $\eta'(x)>0$, and $L(x)$ is increasing. By the second of Eqs.~\eqref{eq:candidate}, $\bar r(\Sigma)$ is the solution of $L(x)=\Sigma$, so that $g(x)$ increases for $x<\bar r(\Sigma)$ and decreases for $x>\bar r(\Sigma)$. Its largest value on $[0,x^\star]$ is $g\big(\bar r(\Sigma)\big)$, and $\eta(x_{\mathrm{f}})\Sigma-I(x_{\mathrm{f}})\le\eta\big(\bar r(\Sigma)\big)\Sigma-I\big(\bar r(\Sigma)\big)$ for every admissible profile.

	The four steps together give, for every admissible profile,
	\begin{equation}
		\begin{split}
			\mathscr{F}[r]&\le \mathscr{F}[r_2]\le\eta(x_{\mathrm{f}})\,\Sigma-I(x_{\mathrm{f}})\\
			&\le\eta\big(\bar r(\Sigma)\big)\,\Sigma-I\big(\bar r(\Sigma)\big)\;,
		\end{split}
		\label{eq:chain}
	\end{equation}
	where the first inequality comes from the first step, the second from the second and third steps, and the third from the fourth step.

	\textit{Fifth step: the candidate reaches the bound.} We now take $r=\bar r$ in Eq.~\eqref{eq:chain}. The candidate is admissible, since it starts at the origin and its slope~\eqref{eq:spiral} lies between $0$ and $1$. It never decreases and stays below $\bar r(\Sigma)<x^\star$, so that the first step leaves it unchanged and $x_{\mathrm{f}}=\bar r(\Sigma)$. By Eq.~\eqref{eq:spiral}, at every $\sigma$ its radial speed is the point of tangency $v_0$ of the second step, with $q=\bar r(\sigma)$ and $x_{\mathrm{f}}=\bar r(\Sigma)$, so that Eq.~\eqref{eq:cs} is an equality and the third step gives $\mathscr{F}[\bar r]=\eta\big(\bar r(\Sigma)\big)\Sigma-I\big(\bar r(\Sigma)\big)$. Finally, $x_{\mathrm{f}}=\bar r(\Sigma)$ is the point where $g(x)$ is largest. Every inequality in Eq.~\eqref{eq:chain} is therefore an equality for the candidate. Each of the two conditions found in the first part makes one inequality an equality: Eq.~\eqref{eq:spiral} the inequality between numbers of the second step, and $L\big(\bar r(\Sigma)\big)=\Sigma$ the choice of the final distance in the fourth step.

	The chain~\eqref{eq:chain} and the fifth step show that the largest value of $\mathscr{F}[r]$ exists, that it is reached by the candidate, and that it equals $\eta\big(\bar r(\Sigma)\big)\Sigma-I\big(\bar r(\Sigma)\big)$. With $\Sigma=K_j(t)$, Eq.~\eqref{eq:compact} then gives
	\begin{align}
		K_{j+1}(t)&\le\Psiq\big(K_j(t)\big)\;,\label{eq:Kbound}\\
		\Psiq\big(K_j(t)\big)&\equiv\max_{0\le x\le x^\star}\big[\eta(x)K_j(t)-I(x)\big]\;,
		\label{eq:Phi}
	\end{align}
	where the maximum is reached at $x=\bar r\big(K_j(t)\big)$, fixed by $L\Big(\bar r\big(K_j(t)\big)\Big)=K_j(t)$.

	\subsection{A qubit Hamiltonian that reaches the bound}
	\label{sec:attain}

	We now show that the bound~\eqref{eq:Kbound} is reached by a qubit Hamiltonian, so that $\Psiq\big(K_j(t)\big)$ is the exact worst case of one step. By Eq.~\eqref{eq:compact}, $K_2(t)$ depends on the Hamiltonian only through the radial profile and the length $K_1(t)$, so that every Hamiltonian whose curve has length $K_1(t)=\Sigma$ and a radial profile that satisfies Eqs.~\eqref{eq:candidate} reaches the bound. An example can be read directly from Eqs.~\eqref{eq:candidate}, because the first of them already contains the two components of a velocity of length one. Squaring it, we obtain
	\begin{equation}
		\big(\dot{\bar r}(\sigma)\big)^2+\left[\frac{\eta\big(\bar r(\sigma)\big)}{\eta\big(\bar r(\Sigma)\big)}\right]^2=1\;.
		\label{eq:unitvelocity}
	\end{equation}

	We therefore take a curve traveled at speed one, $\mathrm{h}_1(t)=1$, so that $K_1(t)=t$ and the length traveled $\sigma$ is the time, with $t$ in $[0,\Sigma]$. By Eq.~\eqref{eq:radial}, the radial part of the velocity is the derivative of the distance from the origin, which is now $\dot{\bar r}(t)$. Since the speed is one, the orthogonal part has length $\sqrt{1-\big(\dot{\bar r}(t)\big)^2}$, which is the square root in Eq.~\eqref{eq:compact}, and by Eq.~\eqref{eq:unitvelocity} with $\sigma=t$ this length is $\eta\big(\bar r(t)\big)/\eta\big(\bar r(\Sigma)\big)$.

	It remains to choose the direction of the orthogonal part. We take the simplest case, a curve in a plane, with an angle $\theta(t)$ such that $\theta(0)=0$,
	\begin{equation}
		\mathbf{f}_1(t)=\bar r(t)\,\big(\cos\theta(t),\sin\theta(t),0\big)\;.
		\label{eq:f1example}
	\end{equation}
	The Hamiltonian is then $H(t)=\mathbf{h}_1(t)\cdot\mathbf{T}$, with
	\begin{equation}
		\begin{split}
			\mathbf{h}_1(t)={}&\dot{\bar r}(t)\,\big(\cos\theta(t),\sin\theta(t),0\big)\\
			&+\frac{\eta\big(\bar r(t)\big)}{\eta\big(\bar r(\Sigma)\big)}\big(-\sin\theta(t),\cos\theta(t),0\big)\;,
		\end{split}
		\label{eq:attain}
	\end{equation}
	\begin{figure}[!t]
		\centering
		\includegraphics[width=\linewidth]{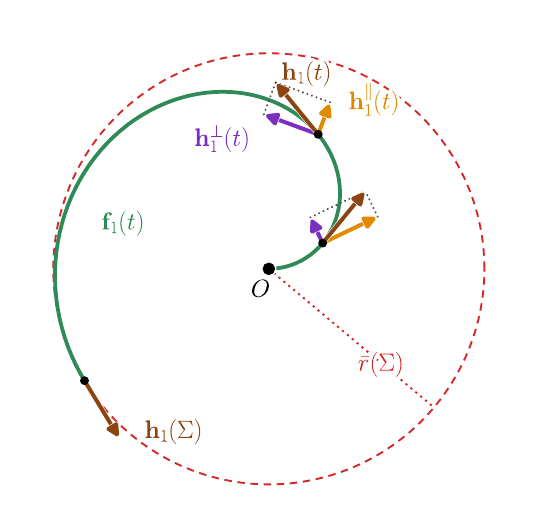}
		\caption{The curve $\mathbf{f}_1(t)$ of the Hamiltonian~\eqref{eq:attain} for $\Sigma=\rhoq$ (green). The curve is numerically plotted as described in Sec.~\ref{sec:attain}. At two points the velocity $\mathbf{h}_1(t)$ (brown) is split into its radial part $\mathbf{h}^{\parallel}_1(t)$ (orange) and its orthogonal part $\mathbf{h}^{\perp}_1(t)$ (violet). The curve ends on the circle of radius $\bar r(\Sigma)$ (red, dashed), where $\mathbf{h}_1(\Sigma)$ has no radial part.}
		\label{fig:example}
	\end{figure}%
	where the first term is the radial part of the velocity and the second is its orthogonal part (Fig.~\ref{fig:example}). The angle $\theta(t)$ is fixed by $\mathbf{h}_1(t)=\dot{\mathbf{f}}_1(t)$, which follows from $\mathbf{f}_1(t)=\int_0^t\rmd s\,\mathbf{h}_1(s)$. The derivative of Eq.~\eqref{eq:f1example} has the same first term as Eq.~\eqref{eq:attain}, and its second term is $\bar r(t)\,\dot\theta(t)\big(-\sin\theta(t),\cos\theta(t),0\big)$. The two second terms coincide if $\bar r(t)\,\dot\theta(t)=\eta\big(\bar r(t)\big)/\eta\big(\bar r(\Sigma)\big)$, that is, if
	\begin{equation}
		\theta(t)=\int_0^t\!\rmd s\;\frac{\eta\big(\bar r(s)\big)}{\eta\big(\bar r(\Sigma)\big)\,\bar r(s)}\;.
		\label{eq:thetaexample}
	\end{equation}
	The integrand is finite also at $s=0$, where it equals $1/\eta\big(\bar r(\Sigma)\big)$, because $\eta(x)/x$ tends to $1$ for small $x$. For Fig.~\ref{fig:example} we set $\Sigma=\rhoq$ and determine the final radius from $L\big(\bar r(\Sigma)\big)=\Sigma$. Since the speed is one, $\sigma=t$: we numerically invert Eq.~\eqref{eq:sigmar} to obtain $\bar r(t)$, then evaluate the integral in Eq.~\eqref{eq:thetaexample} to obtain $\theta(t)$. These two functions give the curve $\mathbf{f}_1(t)$ through Eq.~\eqref{eq:f1example}.

	Equation~\eqref{eq:f1example} shows that the curve is a kind of spiral, since its distance from the origin and its angle both grow along it. By the first of Eqs.~\eqref{eq:candidate}, $\dot{\bar r}(t)>0$ for $t<\Sigma$, because $\bar r(t)<\bar r(\Sigma)$ and $\eta(x)$ increases up to $x^\star$, and by Eq.~\eqref{eq:thetaexample}, $\dot\theta(t)>0$. Dividing the two derivatives, the growth of the distance for a unit of angle is
	\begin{equation}
		\frac{\dot{\bar r}(t)}{\dot\theta(t)}=\frac{\bar r(t)\,\sqrt{\eta^2\big(\bar r(\Sigma)\big)-\eta^2\big(\bar r(t)\big)}}{\eta\big(\bar r(t)\big)}\;,
		\label{eq:spiralslope}
	\end{equation}
	which is positive for $t<\Sigma$ and vanishes at $t=\Sigma$. The curve leaves the origin in the radial direction, since $\eta(0)=0$ makes the orthogonal part vanish, and by Eq.~\eqref{eq:candidateends} it reaches the circle of radius $\bar r(\Sigma)$ tangentially, staying inside it.

	The curve~\eqref{eq:f1example} has length $K_1(\Sigma)=\Sigma$ and radial profile $\bar r(\sigma)$, so that Eq.~\eqref{eq:compact} gives $K_2(\Sigma)=\mathscr{F}[\bar r]$, and by the fifth step of Sec.~\ref{sec:proof} this is $\eta\big(\bar r(\Sigma)\big)\,\Sigma-I\big(\bar r(\Sigma)\big)=\Psiq(\Sigma)$. The unit speed, the plane and the sign of $\dot\theta(t)$ are choices that only serve to make the example explicit.
	

	\section{A sufficient condition for convergence}
	\label{sec:radius}

	\subsection{The fixed point}
	\label{sec:fixedpoint}

	As in Sec.~\ref{sec:fer}, the bound~\eqref{eq:Kbound} makes the expansion converge when $\Psiq\big(K_j(t)\big)<K_j(t)$, that is, when $\eta(x)K_j(t)-I(x)<K_j(t)$ for every $x$ in $[0,x^\star]$. For the values of $x$ with $\eta(x)\le1$ this always holds, because $I(x)>0$. For the values with $\eta(x)>1$ it holds when $K_j(t)<I(x)/[\eta(x)-1]$. A sufficient condition for convergence is therefore $K_j(t)<\rhoq$, where $\rhoq$ is the smallest of these thresholds:
	\begin{equation}
		\rhoq = \min_{x^\circ<x\le x^\star}\;\frac{I(x)}{\eta(x)-1} \approx 6.2106\;,
		\label{eq:rho2}
	\end{equation}
	where $x^\circ$, about $1.1656$ (Fig.~\ref{fig:etaL}), is the point where $\eta(x)$ first reaches $1$. Dividing by $\pi$, we get 
	\begin{equation}
		\frac{\rhoq}{\pi} \approx 1.9769\;.
		\label{eq:rho2pi}
	\end{equation}

	\subsection{Convergence and error bound}
	\label{sec:convergence}

	Every effective Hamiltonian of a qubit is again a qubit Hamiltonian, so the bound $K_{j+1}(t)\le\Psiq\big(K_j(t)\big)$ holds at every step. The map $\Psiq(\Sigma)$ is non-decreasing, as the maps $\Psi_i(x)$: by Eq.~\eqref{eq:Phi} it is the largest of the functions $\eta(x)\Sigma-I(x)$, none of which decreases with $\Sigma$, since $\eta(x)\ge0$. The argument of Sec.~\ref{sec:fer} therefore applies with $\Psiq(\Sigma)$ in place of $\Psi_i(x)$. So far we have considered Hamiltonians $H(t)=\mathbf{h}(t)\cdot\mathbf{T}$. A general qubit Hamiltonian also contains a multiple of the identity, $H(t)=h_0(t)\openone+\mathbf{h}(t)\cdot\mathbf{T}$ with $h_0(t)$ real. This term commutes with everything: it changes only the first factor of the product~\eqref{eq:product}, by the global phase $\exp[-\rmi\int_0^t\rmd s\,h_0(s)]$, and leaves all the effective Hamiltonians $H_j(t)$ with $j\ge2$ unchanged~\cite{bagnasacco2026fer}. We conclude that for every qubit Hamiltonian the Fer expansion converges in operator norm whenever
	\begin{equation}
		\int_0^t\!\rmd s\;\|\mathbf{h}(s)\| < \rhoq\;,
		\label{eq:main}
	\end{equation}
	with the truncation error bounded by
	\begin{equation}
		\Big\|U(t)-\rme^{F_1(t)}\cdots\rme^{F_N(t)}\Big\| \le \Psiq^{\circ N}\Big(\int_0^t\!\rmd s\;\|\mathbf{h}(s)\|\Big)\;.
		\label{eq:mainerror}
	\end{equation}
	Since $\eta'(x)$ decreases from $1$ to $0$ on $[0,x^\star]$, changing variable from $u$ to $\eta(u)$ in $I(x)$ gives $I(x)\ge\pi\eta^2(x)/4$, the area of a quarter disk of radius $\eta(x)$. Therefore $\eta(x)\Sigma-I(x)\le\eta(x)\Sigma-\pi\eta^2(x)/4\le \Sigma^2/\pi$ for every $x$ in $[0,x^\star]$, that is, $\Psiq(\Sigma)\le\Sigma^2/\pi$. Once $K_j(t)$ drops below $\pi$, this gives $K_{j+1}(t)\le K^2_j(t)/\pi$, and the remainder decays doubly exponentially.

	\subsection{Half-width and comparison}
	\label{sec:comparison}

	The eigenvalues of $H(s)$ are $h_0(s)\pm\|\mathbf{h}(s)\|$, so $\|\mathbf{h}(s)\|$ is the half-width of the spectrum, $[\lambda_{\max}(s)-\lambda_{\min}(s)]/2$, and Eq.~\eqref{eq:main} is the condition~\eqref{eq:intro}. In terms of the operator norm, $\|H(s)\|=|h_0(s)|+\|\mathbf{h}(s)\|$. For a traceless Hamiltonian Eq.~\eqref{eq:main} therefore reads $\int_0^t\rmd s\,\|H(s)\|<\rhoq$, and for a general one this condition is sufficient.

	The bound $\rhoq$ should be compared with the values $2$ of Ref.~\cite{blanes1998} and about $2.6058$ of Ref.~\cite{bagnasacco2026fer}, which hold for Hermitian generators of any dimension, and with the radius $\pi$ of the Magnus expansion~\cite{moan2006,casas2007}. An example is the rotating field of Ref.~\cite{lakos2019}, $\mathbf{h}(s)=\gamma\,\big(\cos(2s),0,-\sin(2s)\big)$ with $0\le s\le\pi$ and $\gamma>0$, whose cumulative norm at the final time is $\gamma\pi$. The Magnus series diverges for every $\gamma>1$, that is, as soon as the cumulative norm exceeds $\pi$, while by Eq.~\eqref{eq:main} the Fer expansion converges for every $\gamma<\rhoq/\pi$, about $1.98$. For a qubit, the guaranteed Fer threshold in cumulative norm is thus about $1.98$ times the optimal universal Magnus threshold.

	\section{Conclusions}
	\label{sec:conclusions}

	Our analysis shows how the geometry of qubit dynamics can sharpen the convergence criterion for the Fer expansion. The key is the relation between each effective Hamiltonian and its own time integral: the component that increases the norm of the integral is precisely the component erased by the next step. The norm of the integral and the size of the next step therefore cannot be maximized independently. For a qubit this competition is described exactly by Eq.~\eqref{eq:exact}, which turns the search for the worst case of one step into a variational problem on the radial profile of a curve.

	The resulting one-step bound is reached by a qubit Hamiltonian, so that the convergence threshold obtained here is optimal among those derived by iterating a universal bound based only on the time integral of the norm of each effective Hamiltonian. We have not proved that it coincides with the exact convergence radius of the Fer expansion for a qubit: a larger guaranteed threshold would require information beyond this universal one-step bound.

	Extending this approach to systems with more than two levels requires a richer geometric description and whether this additional information can yield a stronger general convergence guarantee in the operator norm remains open.

	\begin{acknowledgments}
	We thank Kazuya Yuasa (Department of Physics, Waseda University, Tokyo, Japan) and Daniel Burgarth (Department Physik, Friedrich-Alexander-Universit\"at Erlangen--N\"urnberg, Erlangen, Germany), with whom our work on the Fer expansion started, for many discussions on the Fer expansion in general and for the ideas that came out of them.
	\end{acknowledgments}
	
	\section*{Use of AI}

	Claude Code (Anthropic), with the model Claude Opus 5.5, was used under the direction of the authors as a research tool in this work. It played an important role in suggesting mathematical tools for the argument. The interpretation of teh results, their development and all the remaining work are due to the authors. They build on our previous work on the Fer expansion~\cite{bagnasacco2026fer,bagnasacco2026su2}, within a research program on the Fer expansion and related product expansions that both authors have pursued for several years. Claude was also used to edit the manuscript, in the last revision with the model Claude Fable 5.2. The authors checked all derivations, numerical values, and figures, and take full responsibility for the content.

	\makeatletter\@booleantrue\raggedcolumn@sw\makeatother
	\bibliographystyle{prsty-title-hyperref}
	\bibliography{bib_pra}

@Article{kieferova2019,
  author  = {Kieferov{\'a}, M{\'a}ria and Scherer, Artur and Berry, Dominic W.},
  journal = {Phys. Rev. A},
  title   = {Simulating the dynamics of time-dependent {H}amiltonians with a truncated {D}yson series},
  year    = {2019}, volume = {99}, number = {4}, pages = {042314},
  doi     = {10.1103/PhysRevA.99.042314}, url = {https://doi.org/10.1103/PhysRevA.99.042314},
}

@Article{berry2020,
  author  = {Berry, Dominic W. and Childs, Andrew M. and Su, Yuan and Wang, Xin and Wiebe, Nathan},
  journal = {Quantum},
  title   = {Time-dependent {H}amiltonian simulation with {$L^1$}-norm scaling},
  year    = {2020}, volume = {4}, pages = {254},
  doi     = {10.22331/q-2020-04-20-254}, url = {https://doi.org/10.22331/q-2020-04-20-254},
}

@Article{casares2024,
  author  = {Moreno Casares, Pablo Antonio and Zini, Modjtaba Shokrian and Arrazola, Juan Miguel},
  journal = {Quantum},
  title   = {Quantum simulation of time-dependent {H}amiltonians via commutator-free quasi-{M}agnus operators},
  year    = {2024}, volume = {8}, pages = {1567},
  doi     = {10.22331/q-2024-12-17-1567}, url = {https://doi.org/10.22331/q-2024-12-17-1567},
}

@Article{fang2025,
  author  = {Fang, Di and Liu, Diyi and Sarkar, Rahul},
  journal = {Commun. Math. Phys.},
  title   = {Time-dependent {H}amiltonian simulation via {M}agnus expansion: algorithm and superconvergence},
  year    = {2025}, volume = {406}, number = {6}, pages = {128},
  doi     = {10.1007/s00220-025-05314-5}, url = {https://doi.org/10.1007/s00220-025-05314-5},
}

@Article{dyson,
  author    = {Dyson, F. J.},
  journal   = {Phys. Rev.},
  title     = {The $S$ matrix in quantum electrodynamics},
  year      = {1949},
  issn      = {0031-899X},
  month     = jun,
  number    = {11},
  pages     = {1736--1755},
  volume    = {75},
  doi       = {10.1103/physrev.75.1736},
  publisher = {American Physical Society (APS)},
  url       = {https://doi.org/10.1103/PhysRev.75.1736},
}

@Article{magnus,
  author    = {Magnus, Wilhelm},
  journal   = {Commun. Pure Appl. Math.},
  title     = {On the exponential solution of differential equations for a linear operator},
  year      = {1954},
  issn      = {1097-0312},
  month     = nov,
  number    = {4},
  pages     = {649--673},
  volume    = {7},
  doi       = {10.1002/cpa.3160070404},
  publisher = {Wiley},
  url       = {https://doi.org/10.1002/cpa.3160070404},
}

@Article{klarsfeld1989,
  author  = {Klarsfeld, S. and Oteo, J. A.},
  journal = {Phys. Rev. A},
  title   = {Recursive generation of higher-order terms in the {M}agnus expansion},
  year    = {1989},
  volume  = {39}, number = {7}, pages = {3270--3273},
  doi     = {10.1103/PhysRevA.39.3270},
  url     = {https://doi.org/10.1103/PhysRevA.39.3270},
}

@Article{blanes2009,
  author    = {Blanes, S. and Casas, F. and Oteo, J. A. and Ros, J.},
  journal   = {Phys. Rep.},
  title     = {The {Magnus} expansion and some of its applications},
  year      = {2009},
  issn      = {0370-1573},
  month     = jan,
  number    = {5--6},
  pages     = {151--238},
  volume    = {470},
  doi       = {10.1016/j.physrep.2008.11.001},
  publisher = {Elsevier BV},
  url       = {https://doi.org/10.1016/j.physrep.2008.11.001},
}

@Article{moan2006,
  author    = {Moan, Per Christian and Niesen, Jitse},
  journal   = {Found. Comput. Math.},
  title     = {Convergence of the {Magnus} series},
  year      = {2008},
  issn      = {1615-3383},
  month     = jun,
  number    = {3},
  pages     = {291--301},
  volume    = {8},
  doi       = {10.1007/s10208-007-9010-0},
  publisher = {Springer Science and Business Media LLC},
  url       = {https://doi.org/10.1007/s10208-007-9010-0},
}

@Article{casas2007,
  author  = {Casas, Fernando},
  journal = {J. Phys. A: Math. Theor.},
  title   = {Sufficient conditions for the convergence of the {M}agnus expansion},
  year    = {2007},
  volume  = {40}, number = {50}, pages = {15001--15017},
  doi     = {10.1088/1751-8113/40/50/006},
  url     = {https://doi.org/10.1088/1751-8113/40/50/006},
}

@Article{suzuki,
  author    = {Suzuki, Masuo},
  journal   = {Commun. Math. Phys.},
  title     = {Generalized {T}rotter's formula and systematic approximants of exponential operators and inner derivations with applications to many-body problems},
  year      = {1976},
  number    = {2},
  pages     = {183--190},
  volume    = {51},
  doi       = {10.1007/BF01609348},
  url       = {https://doi.org/10.1007/BF01609348},
}

@Article{lloyd1996,
  author = {Lloyd, Seth}, journal = {Science},
  title = {Universal quantum simulators},
  year = {1996}, volume = {273}, number = {5278}, pages = {1073--1078},
  doi = {10.1126/science.273.5278.1073}, url = {https://doi.org/10.1126/science.273.5278.1073},
}

@Article{childs2021,
  author = {Childs, Andrew M. and Su, Yuan and Tran, Minh C. and Wiebe, Nathan and Zhu, Shuchen},
  journal = {Phys. Rev. X}, title = {Theory of {T}rotter error with commutator scaling},
  year = {2021}, volume = {11}, number = {1}, pages = {011020},
  doi = {10.1103/PhysRevX.11.011020}, url = {https://doi.org/10.1103/PhysRevX.11.011020},
}

@Article{casas-zassenhaus,
  author    = {Casas, Fernando and Murua, Ander and Nadinic, Mladen},
  journal   = {Comput. Phys. Commun.},
  title     = {Efficient computation of the {Z}assenhaus formula},
  year      = {2012},
  number    = {11},
  pages     = {2386--2391},
  volume    = {183},
  doi       = {10.1016/j.cpc.2012.06.006},
  url       = {https://doi.org/10.1016/j.cpc.2012.06.006},
}

@Article{wilcox,
  author    = {Wilcox, R. M.},
  journal   = {J. Math. Phys.},
  title     = {Exponential operators and parameter differentiation in quantum physics},
  year      = {1967},
  issn      = {1089-7658},
  month     = apr,
  number    = {4},
  pages     = {962--982},
  volume    = {8},
  doi       = {10.1063/1.1705306},
  publisher = {AIP Publishing},
  url       = {https://doi.org/10.1063/1.1705306},
}

@Article{arnal2021,
  author  = {Arnal, Ana and Casas, Fernando and Chiralt, Cristina and Oteo, Jos{\'e} Angel},
  journal = {Mathematics},
  title   = {A unifying framework for perturbative exponential factorizations},
  year    = {2021},
  volume  = {9},
  number  = {6},
  pages   = {637},
  doi     = {10.3390/math9060637},
  url     = {https://doi.org/10.3390/math9060637},
}

@Article{fer,
  author    = {Fer, Francis},
  journal   = {Bull. Cl. Sci., Acad. R. Belg.},
  title     = {R{\'{e}}solution de l'{\'{e}}quation matricielle $dU/dt = pU$ par produit infini d'exponentielles matricielles},
  year      = {1958},
  issn      = {0001-4141},
  number    = {1},
  pages     = {818--829},
  volume    = {44},
  doi       = {10.3406/barb.1958.68918},
  publisher = {PERSEE Program},
  url       = {https://doi.org/10.3406/barb.1958.68918},
}

@Article{blanes1998,
  author  = {Blanes, S. and Casas, F. and Oteo, J. A. and Ros, J.},
  journal = {J. Phys. A: Math. Gen.},
  title   = {{M}agnus and {F}er expansions for matrix differential equations: the convergence problem},
  year    = {1998},
  volume  = {31}, number = {1}, pages = {259--268},
  doi     = {10.1088/0305-4470/31/1/023},
  url     = {https://doi.org/10.1088/0305-4470/31/1/023},
}

@Article{iserles1984,
  author  = {Iserles, Arieh},
  journal = {Numer. Math.},
  title   = {Solving linear ordinary differential equations by exponentials of iterated commutators},
  year    = {1984},
  volume  = {45},
  number  = {2},
  pages   = {183--199},
  doi     = {10.1007/BF01389464},
  url     = {https://doi.org/10.1007/BF01389464},
}

@Article{iserles2000,
  author  = {Iserles, Arieh and Munthe-Kaas, Hans Z. and N{\o}rsett, Syvert P. and Zanna, Antonella},
  journal = {Acta Numer.},
  title   = {{L}ie-group methods},
  year    = {2000},
  volume  = {9}, pages = {215--365},
  doi     = {10.1017/S0962492900002154},
  url     = {https://doi.org/10.1017/S0962492900002154},
}

@Article{zanna1999,
  author  = {Zanna, A.},
  journal = {SIAM J. Numer. Anal.},
  title   = {Collocation and relaxed collocation for the {F}er and the {M}agnus expansions},
  year    = {1999},
  volume  = {36}, number = {4}, pages = {1145--1182},
  doi     = {10.1137/S0036142997326616},
  url     = {https://doi.org/10.1137/S0036142997326616},
}

@Article{madhu2006,
  author  = {Madhu, P. K. and Kurur, Narayanan D.},
  journal = {Chem. Phys. Lett.},
  title   = {{F}er expansion for effective propagators and {H}amiltonians in {NMR}},
  year    = {2006},
  volume  = {418},
  number  = {1--3},
  pages   = {235--238},
  doi     = {10.1016/j.cplett.2005.10.134},
  url     = {https://doi.org/10.1016/j.cplett.2005.10.134},
}

@Article{mananga2016,
  author  = {Mananga, Eugene Stephane},
  journal = {Phys. Rep.},
  title   = {On the {F}er expansion: Applications in solid-state nuclear magnetic resonance and physics},
  year    = {2016},
  volume  = {608},
  pages   = {1--41},
  doi     = {10.1016/j.physrep.2015.10.006},
  url     = {https://doi.org/10.1016/j.physrep.2015.10.006},
}

@Article{haeberlen1968,
  author = {Haeberlen, U. and Waugh, J. S.}, journal = {Phys. Rev.},
  title = {Coherent averaging effects in magnetic resonance},
  year = {1968}, volume = {175}, number = {2}, pages = {453--467},
  doi = {10.1103/PhysRev.175.453}, url = {https://doi.org/10.1103/PhysRev.175.453},
}

@Article{mananga2011,
  author = {Mananga, Eug\`{e}ne S. and Charpentier, Thibault}, journal = {J. Chem. Phys.},
  title = {Introduction of the {F}loquet-{M}agnus expansion in solid-state nuclear magnetic resonance spectroscopy},
  year = {2011}, volume = {135}, number = {4}, pages = {044109},
  doi = {10.1063/1.3610943}, url = {https://doi.org/10.1063/1.3610943},
}

@Article{viola1999,
  author = {Viola, Lorenza and Knill, Emanuel and Lloyd, Seth}, journal = {Phys. Rev. Lett.},
  title = {Dynamical decoupling of open quantum systems},
  year = {1999}, volume = {82}, number = {12}, pages = {2417--2421},
  doi = {10.1103/PhysRevLett.82.2417}, url = {https://doi.org/10.1103/PhysRevLett.82.2417},
}

@Article{casas1996,
  author  = {Casas, Fernando},
  journal = {Numer. Math.},
  title   = {{F}er's factorization as a symplectic integrator},
  year    = {1996},
  volume  = {74},
  number  = {3},
  pages   = {283--303},
  doi     = {10.1007/s002110050217},
  url     = {https://doi.org/10.1007/s002110050217},
}

@Article{zanna2001,
  author  = {Zanna, Antonella},
  journal = {Appl. Numer. Math.},
  title   = {The {F}er expansion and time-symmetry: a {S}trang-type approach},
  year    = {2001},
  volume  = {39},
  number  = {3--4},
  pages   = {435--459},
  doi     = {10.1016/S0168-9274(01)00085-X},
  url     = {https://doi.org/10.1016/S0168-9274(01)00085-X},
}

@Article{stampfli1970,
  author  = {Stampfli, Joseph G.},
  journal = {Pac. J. Math.},
  title   = {The norm of a derivation},
  year    = {1970}, volume = {33}, number = {3}, pages = {737--747},
  doi     = {10.2140/pjm.1970.33.737}, url = {https://doi.org/10.2140/pjm.1970.33.737},
}

@Misc{lakos2019,
  author       = {Lakos, Gyula},
  title        = {Convergence estimates for the {M}agnus expansion {II}. {$C^*$}-algebras},
  year         = {2019},
  howpublished = {\href{https://arxiv.org/abs/1910.03328}{arXiv:1910.03328}},
}

@Article{feynman,
  author    = {Feynman, Richard P. and Vernon, Frank L. and Hellwarth, Robert W.},
  journal   = {J. Appl. Phys.},
  title     = {Geometrical representation of the Schr{\"{o}}dinger equation for solving maser problems},
  year      = {1957},
  issn      = {1089-7550},
  month     = jan,
  number    = {1},
  pages     = {49--52},
  volume    = {28},
  doi       = {10.1063/1.1722572},
  publisher = {AIP Publishing},
  url       = {https://doi.org/10.1063/1.1722572},
}

@Article{Vandersypen2005,
  author    = {Vandersypen, Lieven M. K. and Chuang, Isaac L.},
  journal   = {Rev. Mod. Phys.},
  title     = {{NMR} techniques for quantum control and computation},
  year      = {2005},
  month     = jan,
  number    = {4},
  pages     = {1037--1069},
  volume    = {76},
  doi       = {10.1103/RevModPhys.76.1037},
  publisher = {American Physical Society},
  url       = {https://doi.org/10.1103/RevModPhys.76.1037},
}

@Article{bloch,
  author    = {Bloch, F.},
  journal   = {Phys. Rev.},
  title     = {Nuclear induction},
  year      = {1946},
  issn      = {0031-899X},
  month     = oct,
  number    = {7--8},
  pages     = {460--474},
  volume    = {70},
  doi       = {10.1103/physrev.70.460},
  publisher = {American Physical Society (APS)},
  url       = {https://doi.org/10.1103/PhysRev.70.460},
}

@Misc{bagnasacco2026fer,
  author       = {Bagnasacco, Lorenzo and Giovannetti, Vittorio},
  title        = {Improved convergence radius of the {Fer} expansion for {Hermitian} generators},
  year         = {2026},
  howpublished = {\href{https://arxiv.org/abs/2609.28302}{arXiv:2609.28302}},
}

@Misc{bagnasacco2026su2,
  author       = {Bagnasacco, Lorenzo and Burgarth, Daniel and Giovannetti, Vittorio and Yuasa, Kazuya},
  title        = {The {Fer} expansion of a driven spin under $\mathfrak{su}(2)$ operations cannot terminate beyond first order},
  year         = {2026},
  howpublished = {\href{https://arxiv.org/abs/2610.27979}{arXiv:2610.27979}},
}

@Book{sakurai,
  author    = {Sakurai, J. J. and Napolitano, Jim},
  publisher = {Cambridge University Press},
  title     = {Modern Quantum Mechanics},
  year      = {2020},
  address   = {Cambridge},
  edition   = {3rd},
  doi       = {10.1017/9781108587280},
  url       = {https://doi.org/10.1017/9781108587280},
}

@Article{spinscattering,
  author    = {Bagnasacco, Lorenzo and Taddei, Fabio and Giovannetti, Vittorio},
  journal   = {J. Phys. A: Math. Theor.},
  title     = {Formal integration of electron scattering processes via separation of dynamical and geometric contributions},
  year      = {2025},
  issn      = {1751-8121},
  month     = dec,
  number    = {50},
  pages     = {505302},
  volume    = {58},
  doi       = {10.1088/1751-8121/ae2995},
  publisher = {IOP Publishing},
  url       = {https://doi.org/10.1088/1751-8121/ae2995},
}

@Book{gelfand1963,
  author    = {Gelfand, I. M. and Fomin, S. V.},
  publisher = {Prentice-Hall},
  title     = {Calculus of Variations},
  year      = {1963},
  address   = {Englewood Cliffs, NJ},
  note      = {reprinted by Dover, Mineola, NY, 2000},
  url       = {https://books.google.com/books?vid=ISBN9780486414485},
}

@Book{buttazzo1998,
  author    = {Buttazzo, Giuseppe and Giaquinta, Mariano and Hildebrandt, Stefan},
  publisher = {Clarendon Press},
  title     = {One-dimensional Variational Problems: An Introduction},
  year      = {1998},
  address   = {Oxford},
  url       = {https://books.google.com/books?vid=ISBN9780198504658},
}

@Book{gobbino2016,
  author    = {Gobbino, Massimo},
  publisher = {Universit\`a di Pisa},
  title     = {Corso di {E}lementi di {C}alcolo delle {V}ariazioni: {S}tampato integrale delle lezioni, {A.A.} 2015/2016},
  year      = {2016},
  address   = {Pisa},
  note      = {lecture notes in Italian},
  url       = {https://pagine.dm.unipi.it/gobbino/Home_Page/Files/HP_AD/CdV_16_Book.pdf},
}

@Article{oreshkov2009,
  author  = {Oreshkov, Ognyan and Brun, Todd A. and Lidar, Daniel A.},
  title   = {Scheme for fault-tolerant holonomic computation on stabilizer codes},
  journal = {Phys. Rev. A},
  year    = {2009}, volume = {80}, pages = {022325},
  doi     = {10.1103/PhysRevA.80.022325}, url = {https://doi.org/10.1103/PhysRevA.80.022325},
}

@Article{sjoqvist2012,
  author  = {Sj\"oqvist, Erik and Tong, D. M. and {Mauritz Andersson}, L. and Hessmo, Bj\"orn and Johansson, Markus and Singh, Kuldip},
  title   = {Non-adiabatic holonomic quantum computation},
  journal = {New J. Phys.},
  year    = {2012}, volume = {14}, pages = {103035},
  doi     = {10.1088/1367-2630/14/10/103035}, url = {https://doi.org/10.1088/1367-2630/14/10/103035},
}

\end{document}